\documentclass[conference]{IEEEtran}

\usepackage[utf8]{inputenc} 
\usepackage[T1]{fontenc}
\usepackage{url}
\usepackage{ifthen}
\usepackage{cite}
\usepackage[cmex10]{amsmath} %
\usepackage{mathtools}
\usepackage{booktabs}
\usepackage{amsfonts}
\usepackage{thmtools}

\usepackage[
	ruled,
	vlined,
	boxed, 
	linesnumbered, 
	commentsnumbered]{algorithm2e}

\usepackage{siunitx}
\usepackage{placeins}
\usepackage{tikz}
\usepackage{pgfplots}
\usepackage{comment}
\usepackage{acronym}
\usepackage{makecell}

\definecolor{A1}{RGB}{146,84,3}%
\definecolor{A2}{RGB}{8,51,110}%
\definecolor{A3}{rgb}{0,.59,.51}%
\definecolor{A4}{rgb}{.63,.13,.13}%
\definecolor{A5}{RGB}{160,0,120}%

\usepackage{bm}
\renewcommand{\vec}[1]{\bm{#1}}

\newcommand{\me}{\mathrm{e}}

\usepackage{bm}
\renewcommand{\vec}[1]{\bm{#1}}
\usepackage{bbm}

\newcommand{\trans}[1]{#1^{\mathsf{T}}}

\newcommand{\lto}{\leftarrow}

\newcommand\blfootnote[1]{%
  \begingroup
  \renewcommand\thefootnote{}\footnote{#1}%
  \addtocounter{footnote}{-1}%
  \endgroup
}

\binoppenalty=\maxdimen
\relpenalty=\maxdimen

\acrodef{BCH}{Bose--Chaudhuri--Hocquenghem}
\acrodef{FER}{frame error rate}
\acrodef{BSC}{binary symmetric channel}
\acrodef{DE}{density evolution}
\acrodef{LDPC}{low-density parity-check}
\acrodef{QLDPC}{quantum low-density parity-check}
\acrodef{VN}{variable node}
\acrodef{CN}{check node}
\acrodef{BP}{belief propagation}
\acrodef{NBP}{neural belief propagation}
\acrodef{GB}{generalized bicycle}
\acrodef{PCM}{parity check matrix}
\acrodefplural{PCM}[PCMs]{parity check matrices}
\acrodef{OSD}{ordered statistics decoding}
\acrodef{QEC}{quantum error correction}
\acrodef{QSC}{quantum stabilizer code}
\acrodef{CSS}{Calderbank–Shor–Steane}
\acrodef{NN}{neural network}
\acrodef{LLR}{log-likelihood ratio}
\pgfplotsset{compat=1.16}

\begin{document}

\def\lsbelowcaption{1.5ex}
\title{Neural Belief Propagation Decoding of Quantum LDPC Codes Using Overcomplete Check Matrices}
 \author{%
   \IEEEauthorblockN{Sisi Miao\textsuperscript{\dag}, Alexander Schnerring\textsuperscript{\ddag}, Haizheng Li\textsuperscript{\dag}, and Laurent Schmalen\textsuperscript{\dag}}
   \IEEEauthorblockA{\textsuperscript{\dag}Karlsruhe Institute of Technology (KIT),
Communications Engineering Lab (CEL),
76187 Karlsruhe, Germany\\
\textsuperscript{\ddag}\emph{now with} German Aerospace Center (DLR), Institute of Solar Research, 04001 Almería, Spain\\ E-Mail: \texttt{sisi.miao@kit.edu}}}

\maketitle

\begin{abstract}
    The recent success in constructing asymptotically good quantum low-density parity-check (QLDPC) codes makes this family of codes a promising candidate for error-correcting schemes in quantum computing. However, conventional belief propagation (BP) decoding of QLDPC codes does not yield satisfying performance due to the presence of unavoidable short cycles in their Tanner graph and the special degeneracy phenomenon. In this work, we propose to decode QLDPC codes based on a check matrix with redundant rows, generated from linear combinations of the rows in the original check matrix. This approach yields a significant improvement in decoding performance with the additional advantage of very low decoding latency. Furthermore, we propose a novel neural belief propagation decoder based on the quaternary BP decoder of QLDPC codes which leads to further decoding performance improvements.
\end{abstract}

\section{Introduction}
\blfootnote{This work has received funding from the European Research Council (ERC)
    under the European Union’s Horizon 2020 research and innovation programme
    (grant agreement No. 101001899).}
\Ac{QEC} is an essential part of fault-tolerant quantum computing. Recent breakthroughs in designing asymptotically good \ac{QLDPC} codes with non-vanishing rate and minimal distance growing linearly with the code length\cite{tillich2013quantum,panteleev2021quantum,panteleev2022asymptotically,breuckmann21balanced} make \ac{QLDPC} codes a promising candidate for future \ac{QEC} schemes. In classical coding, \ac{LDPC} codes are typically decoded with a message-passing decoder such as the \ac{BP} decoder, which usually works well when the Tanner graph has a girth of at least $6$. However, this condition cannot be fulfilled in the case of \ac{QLDPC} codes, where $4$-cycles cannot be avoided by construction. Therefore, to achieve good decoding performance for \ac{QLDPC} codes similar to their classical counterparts, it is crucial to modify the \ac{BP} decoder such that short cycles can be tolerated.

A variety of methods have been proposed to solve this issue. These can be grouped into two categories. The first category contains approaches that modify the \ac{BP} decoder itself, for example, message normalization and offsets~\cite{kuo2020refined,lai2021log}, layered scheduling\cite{panteleev2021degenerate,raveendran2021trapping}, and matrix augmentation\cite{rigby2019modified}. The second category contains methods that apply post-processing to the \ac{BP} decoder output, e.g., \ac{OSD}\cite{panteleev2021degenerate,roffe2020decoding}, random perturbation~\cite{poulin2008iterative}, enhanced feedback~\cite{wang2012enhancedfeedback}, and stabilizer inactivation~\cite{crest2022stabilizer}. However, most of these methods introduce additional decoding latency, which makes them less appealing in \ac{QEC} where decoding has to be performed with ultra low latency.

\Ac{NBP} has been shown to be effective in improving the decoding performance of classical linear codes without increasing the decoding latency, see, e.g.,~\cite{nachmani2016learning,nachmani2018deep,lian2019learned}. Using \ac{NBP} to enhance the decoding of quantum stabilizer codes has been investigated in~\cite{liu2019neural,xiao2019neural} for the binary \ac{BP} decoder (referred to as BP2 decoder), which decodes $\vec{X}$ and $\vec{Z}$ errors separately. Binary decoding is sub-optimal and has a higher error floor compared with a quaternary \ac{BP} decoder  (referred to as BP4 decoder), which takes the correlation between $\vec{X}$ and $\vec{Z}$ errors into account~\cite{panteleev2021degenerate,lai2021log}. The disadvantage of the conventional BP4 decoder is the high complexity due to the passing of vector messages instead of scalar messages as in BP2. The problem is solved by the recently proposed refined BP4 decoder with scalar messages~\cite{kuo2020refined,lai2021log}.

In this paper, we enhance the refined BP4 decoder with \ac{NBP} for QLDPC codes. To avoid typical problems such as vanishing gradients in the training of very deep \acp{NN}, our proposed NBP model is based on an overcomplete check matrix with redundant rows. It is inspired by the observation in classical coding theory that using an overcomplete \ac{PCM} enables more node updates in parallel, which reduces the required number of decoding iterations and combats the effects of short cycles. Consequently, the trained \ac{NBP} network effectively improves the decoding performance by orders of magnitude compared to conventional BP decoding for the considered codes. Moreover, the required number of decoding iterations is reduced significantly, yielding a low-latency decoder.

\section{Preliminaries}
\subsection{Stabilizer Formalism}
\Acp{QSC}~\cite{gottesman1997stabilizer,Calderbank1998quantum} are the quantum analogs of classical linear codes. To define a \ac{QSC}, we first need to define the Pauli operators. For simplicity, we ignore the global phase and consider the $n$-qubit Pauli group $\mathcal{G}_n$ consisting of Pauli operators on $n$ qubits $\vec{\mathcal{P}} = \mathcal{P}_1\otimes\mathcal{P}_2 \otimes \cdots \otimes \mathcal{P}_n$ where $\mathcal{P}_i\in \{\vec{I},\vec{X},\vec{Y},\vec{Z}\}$ are Pauli operators on the $i$-th qubit. Without the risk of confusion, the tensor product $\otimes$ and the identity operator $\vec{I}$ can be omitted. The weight $w$ of a Pauli operator $\vec{\mathcal{P}}$ is the number of non-identity components in the tensor product. For the depolarizing channel with depolarizing probability $\epsilon$ considered in this work, the errors $\vec{X},\vec{Z}$ and $\vec{Y}$ occur equally likely with probability $\frac{\epsilon}{3}$. 
To find a valid stabilizer code, it is crucial to find the stabilizer group $\mathcal{S}$, which is an Abelian subgroup of $\mathcal{G}_n$. This can be transferred into a classical coding problem using the mapping of Pauli errors to binary strings $\vec{\mathcal{P}}\mapsto \begin{pmatrix}x_1\;\cdots\;x_n\mid z_1\;\cdots\;z_n\end{pmatrix}=:\begin{pmatrix}\vec{p}_{X}\mid \vec{p}_{Z}\end{pmatrix}$ with $\mathcal{P}_i = \vec{X}^{x_i}\vec{Z}^{z_i}$.
We can check that Pauli errors $\mathcal{A}$ and $\mathcal{B}$ commute if the symplectic product of their corresponding binary strings $(\vec{a}_X \mid \vec{a}_Z)$ and $(\vec{b}_X \mid \vec{b}_Z)$ is 0, i.e.
\begin{equation}
\sum_{i=1}^{n} a_{X,i}\cdot b_{Z,i}  + \sum_{i=1}^{n} a_{Z,i}\cdot b_{X,i}= 0.
\label{eq:commute_binary}
\end{equation}
Therefore, a stabilizer code can be constructed from a $[2n,k]$ classical binary code given by its full rank \ac{PCM} $\vec{H} = \begin{pmatrix}
\vec{H}_X \mid \vec{H}_Z\\
\end{pmatrix}$ of size $m=2n-k$ by $2n$ which fulfills the symplectic criterion:
\begin{equation}
\vec{H}_X \trans{\vec{H}_Z} + \vec{H}_Z\trans{\vec{H}_X} =  \vec{0}.
\label{eq:symplectic criterion}
\end{equation}
\ac{CSS} codes are a special kind of stabilizer codes where 
\[\vec{H} =\left(\begin{array}{c}
\vec{H}_X'    \\
\vec{0}
\end{array}
\middle\vert
\begin{array}{c}
\vec{0} \\  \vec{H}_Z'
\end{array}
\right).\]
In this case, \eqref{eq:symplectic criterion} holds if $\vec{H}_X' \vec{H}_Z^{'\mathsf{T}} = \vec{0}$ and the code construction can be simplified. \Ac{QLDPC} codes are defined as a family of stabilizer codes whose row and column weights are upper bounded by a relatively small constant independent of the block length, i.e., $\vec{H}$ is sparse. Most of the successful \ac{QLDPC} codes constructed so far are \ac{CSS} codes and the \ac{QLDPC} codes considered in this paper are also \ac{CSS} codes.

To decode the four types of Pauli errors jointly, it is convenient to consider the quaternary form of $\vec{H}$, denoted as $\vec{S}\in\text{GF(4)}^{m\times n}$, where GF(4) consist of the elements $\{0,1,\omega,\bar{\omega}\}$. $\vec{H}$ can be converted to $\vec{S}$ using the mapping of binary strings to strings over GF(4) $\begin{pmatrix}x_1\;\cdots\;x_n\mid z_1\;\cdots\;z_n\end{pmatrix}\mapsto \begin{pmatrix}p_1\;\cdots\;p_n\end{pmatrix}=:\vec{p}$ with $p_i =x_i \omega  +  z_i \bar{\omega}$. Then we can check that for $\mathcal{A}$ and $\mathcal{B}$, mapped to vectors $\vec{a}$ and $\vec{b}$ over GF(4), \eqref{eq:commute_binary} is equivalent to  $\langle \vec{a},\vec{b} \rangle = \sum_{i=1}^{n}\langle a_{i}, b_{i}\rangle = 0$, where $\langle\cdot,\cdot \rangle$ denotes the \emph{trace inner product} over GF(4). 

The matrix $\vec{S}$ is called the \emph{check matrix} and every row of $\vec{S}$ is called a \emph{check}. The checks correspond to the $m$ stabilizers which generate the stabilizer group $\mathcal{S}$. An $[[n,k,d]]$ stabilizer code is a $2^k$-dimensional subspace of the $n$-qubit Hilbert space $(\mathbb{C}^2)^{\otimes n}$ defined as the common +1 eigenspace of $\mathcal{S}$. Additionally, we define $\mathcal{N}(\mathcal{S})$ to be the normalizer of $\mathcal{S}$ and $\vec{S}^{\perp}$ to be the matrix containing the vectors corresponding to the $2n-m$ generators of $\mathcal{N}(\mathcal{S})$. The minimum distance $d$ is defined as the lowest error weight in $\mathcal{N}(\mathcal{S})\backslash \mathcal{S}$. A code is called \emph{degenerate} if $\mathcal{S}$ contains errors of weight less than $d$.

\subsection{Syndrome Decoding of Stabilizer codes}
Let $\vec{\mathcal{E}}\in \mathcal{G}_n$ be an occurred error and $\vec{e}$ be the corresponding error vector over GF(4). To obtain an error syndrome, measurements are performed on the $m$ stabilizer generators. The results are mapped to 0 or 1 indicating whether the error commutes with the corresponding stabilizer generator or not, i.e, the error syndrome is $\vec{z}=(z_1,z_2,\ldots, z_m)\in \{0,1\}^m$, where $z_i = \langle \vec{e}, \vec{S}_i \rangle$ and $\vec{S}_i$ is the $i$-th row of $\vec{S}$.

Let $\hat{\vec{e}}$ be the estimate of $\vec{e}$ by the decoder and $\hat{\vec{\mathcal{E}}}$ be its corresponding Pauli error. The decoder aims to find an $\hat{\vec{e}}$ which yields the same syndrome $\vec{z}$ and such that $\hat{\vec{\mathcal{E}}}\vec{\mathcal{E}}\in \mathcal{S}$. The latter can be checked by
\begin{equation}
\label{eq:correction}
\langle (\vec{e}+\hat{\vec{e}}), \vec{S}_i^{\perp}\rangle=0
\end{equation}
for every row $i\in \{1,2,\ldots, 2n-m\}$ of $\vec{S}^{\perp}$. Two types of decoding failures may happen: One is when the decoder fails to find any $\hat{\vec{e}}$ which matches the syndrome $\vec{z}$. Another is when an inferred error $\hat{\vec{e}}$ is found which matches the syndrome $\vec{z}$ but \eqref{eq:correction} does not hold. This is called an ``unflagged-error'' \cite{liu2019neural} and it leads to an undetectable erroneous state.

\subsection{Belief Propagation Decoder}
The task of inferring an error from a given syndrome for a \ac{QLDPC} code can be carried out with a BP4 decoder.  We use the log-domain refined BP4 decoder proposed in~\cite{lai2021log}. It exploits the fact that the syndrome of a stabilizer code is binary, indicating whether the error commutes with the stabilizer or not, which enables scalar message passing. We briefly review the algorithm here. 

For every \ac{VN} $\mathsf{v}_i$, where $ i\in$ $\{1,2,\ldots, n\}$, we initialize the \ac{LLR} vector $\vec{\Gamma}_{i\to j}$ as $\vec{\Lambda}_i=\begin{pmatrix}
\Lambda_i^{(1)}&\Lambda_i^{(\omega)}&\Lambda_i^{(\bar{\omega})}\\
\end{pmatrix}\in \mathbb{R}^3$ with
\[
\Lambda_i ^{(\zeta)} = \ln \left(\frac{P(e_i=0)}{P(e_i=\zeta)}\right)=\ln \left(\frac{1-\epsilon_0}{\frac{\epsilon_0}{3}}\right)
\]
where $\zeta\in$ GF(4)$\backslash \{0\}$ and $\epsilon_0$ is the estimated physical error probability of the channel. To exchange scalar messages, a \textit{belief-quantization operator} $\lambda_{\eta}: \mathbb{R}^3 \to \mathbb{R}$ is defined as
\[\lambda_{\eta}(\vec{\Lambda}_i) = \ln \left(\frac{P(\langle e_i, \eta\rangle=0)}{P(\langle e_i, \eta\rangle=1)}\right)
= \ln  \left(\frac{1+\me^{-\Lambda^{(\eta)}_i}}{\sum_{\zeta\neq 0, \zeta \neq \eta}\me^{-\Lambda^{(\zeta)}_i}}\right).\]
The operator $\lambda_{\eta}$ turns the \ac{LLR} vector into a scalar \ac{LLR} of the binary random variable $\langle e_i,\eta \rangle$ where $\eta$ runs over the nonzero entries of $\vec{S}$. The initial scalar \ac{VN} messages are calculated as 
\begin{equation}
\lambda_{i\to j} :=\lambda_{S_{ji}}(\vec{\Gamma}_{i\to j})
\label{eq:VNinitial}
\end{equation}
and are passed to the neighboring \acp{CN}.

The outgoing messages of \ac{CN} $\mathsf{c}_j$, $j\in \{1,2,\ldots, m\}$,  are calculated using
\begin{equation}
\label{eq:CNupdate}
\Delta_{i\lto j} = (-1)^{z_j}\cdot 2\tanh^{-1} \left(\prod_{i'\in \mathcal{N}(j)\backslash{i}}\tanh \frac{\lambda_{i'\to j}}{2}\right)
\end{equation} where $\mathcal{N}(j)$ denotes the neighboring VNs of $\mathsf{c}_j$. 

At the \ac{VN} update, we first calculate the \ac{LLR} vector $\vec{\Gamma}_{i\to j}=\begin{pmatrix}
{\Gamma}_{i\to j}^{(1)}&{\Gamma}_{i\to j}^{(\omega)}&{\Gamma}_{i\to j}^{(\bar{\omega})}\\
\end{pmatrix}$ with
\begin{equation}
\label{eq:VNupdate}
\Gamma^{(\zeta)}_{i\to j} = \Lambda_i^{(\zeta)} + \sum_{\substack{j'\in \mathcal{M}(i)\backslash{j}, \\ \langle \zeta, S_{j'i}\rangle=1}}\Delta_{i\lto j'},
\end{equation}
for all $\zeta\in$ GF(4)$\backslash \{0\}$ with $\mathcal{M}(i)$ denoting the neighboring CNs of $\mathsf{v}_i$. Then the outgoing messages $\lambda_{i\to j} = \lambda_{S_{ji}}(\vec{\Gamma}_{i\to j})$ are calculated and passed to the neighboring CNs. 

To estimate the error, a hard decision is performed at the \ac{VN}s by calculating $\vec{\Gamma}_i$ for $i\in \{1,2,\ldots, n\}$ with

\begin{equation}
\label{eq:hard-decision}
\Gamma^{(\zeta)}_{i} = \Lambda_i^{(\zeta)} + \sum_{\substack{j\in \mathcal{M}(i)
        \\ \langle \zeta, S_{ji}\rangle=1}} \Delta_{i\lto j},
\end{equation}
for all $\zeta\in$ GF(4)$\backslash \{0\}$. If all $\Gamma^{(\zeta)}_{i}>0$, then $\hat{e}_i=0$, otherwise $\hat{e}_i=\text{argmin}_{\zeta} \Gamma^{(\zeta)}_{i}$.

The iterative process is performed until the maximum number of iterations $L$ is reached or the syndrome is matched.

\section{Check matrix With Redundant Rows}
\label{sec:soc}
To construct an overcomplete check matrix $\vec{S}_{\text{oc}}$ with redundant rows, we treat the two binary matrices $\vec{H}_X$ and $\vec{H}_Z$ separately and consider them as classical binary \acp{PCM}. For both, $\vec{H}_X$ and $\vec{H}_Z$, redundant rows are generated by linear combinations of the rows of the original matrix. We try to keep the sparsity of the \ac{PCM} by only generating low-weight rows. The task is equivalent to generating low-weight dual codewords in classical coding theory and we use Leon's probabilistic algorithm~\cite{Leon88}. \Ac{BP} decoding is performed on the Tanner graph associated with the overcomplete $\vec{S}_{\text{oc}}$.

It is important to note that using this method does not require additional syndrome measurements which would be costly. Let $\vec{H}$ be either $\vec{H}_X$ or $\vec{H}_Z$. The \ac{PCM} $\vec{H_{\text{oc}}}$ with redundant rows is obtained by $\vec{H_{\text{oc}}} = \vec{MH}$ with $\vec{M}$ being a binary matrix of size $m_{\text{oc}}\times m$. 
The original syndrome is calculated as $\vec{H}\trans{\vec{e}} = \vec{z}$. The new syndrome associated with $\vec{H}_{\text{oc}}$ is given by
\[\vec{z}_{\text{oc}} = \vec{H}_{\text{oc}}\trans{\vec{e}} =\vec{MH}\trans{\vec{e}} = \vec{M}\vec{z}, \]
being a linear mapping of $\vec{z}$ using $\vec{M}$.

The idea of performing \ac{BP} decoding over an overcomplete parity-check matrix has been investigated in~\cite{lian2019learned,halford2006random,bossert1986hard,kothiyal2005iterative,Jiang2006iterative,buchberger2020pruning} for classical linear codes. One of the initial motivations is to perform more node updates in parallel to reduce the effect of short cycles. For \ac{QLDPC} codes, this approach resembles matrix augmentation~\cite{rigby2019modified}, where a fraction of the rows of the check matrix are duplicated. The advantage is that the messages associated with the duplicated check node are magnified which helps in breaking the symmetry during decoding and leading to a (hopefully) correct error estimation~\cite{rigby2019modified}. We demonstrate the extra benefit of the proposed method with a toy example. Consider the $[[7,1,3]]$ quantum \ac{BCH} code with both its $\vec{H}_X'$ and $\vec{H}_Z'$ being the \ac{PCM} of a $[7,4,3]$ \ac{BCH} code:
\[
\Vec{H}_{\text{BCH}} = \begin{pmatrix}
1&0&1&0&1&0&1\\
0&1&1&0&0&1&1\\
0&0&0&1&1&1&1\\
\end{pmatrix}.
\]
Consider the error $\vec{\mathcal{E}} = \vec{Y}_7$. The syndrome $\vec{z}$ of $\vec{e}$ is
\[
\vec{z} = \begin{pmatrix}
1\;1\;1\;1\;1\;1\\
\end{pmatrix}.
\]
Assume that we initialize the decoder with $\epsilon_0=0.1$. According to \eqref{eq:VNinitial}, all the initial \ac{VN} messages are $2.64$. Then, in the first \ac{CN} update, all \ac{CN} messages are $-1.55$ according to \eqref{eq:CNupdate}. Based on these messages, in the next hard decision step, the decoder estimates the error as $\hat{\vec{\mathcal{E}}}=\vec{Y}_3\vec{Y}_5\vec{Y}_6\vec{Y}_7$ which produces the same syndrome as $\vec{z}$. However, \eqref{eq:correction} does not hold and  we end up with an unflagged-error.

\begin{figure}[tb]
    \centering
     \begin{tikzpicture}
       \pgfplotsset{grid style={gray!40}}
       \pgfplotsset{every tick label/.append style={font=\footnotesize}}
        \begin{axis}[%
        name=ax1,
        width=8.2cm,
        height=5.2cm,
        xmin=0.01,
        xmax=0.1,
        xmode=log,
        ymode=log,
        ymin=1e-3,
        ymax=0.2,
        yminorticks,
        axis background/.style={fill=white, mark size=1.5pt},
        xmajorgrids,
        xminorgrids,
        ymajorgrids,
        yminorgrids,
        x tick label style={/pgf/number format/fixed},
        minor x tick num=1,
        ylabel={FER},
        xlabel={Physical error rate $\epsilon$},
        label style={font=\small},
        legend cell align={left},
        legend style={at={(1,0)},anchor=south east,fill opacity=1, text opacity = 1,legend columns=1, row sep = 0pt,font=\footnotesize}
]

   \addplot [color=A5!40, dotted, line width=1.5pt, mark=+, mark options={solid, A5!40, fill=white, mark size=3.5pt}]table[row sep=crcr]{%
0.1  0.16755208333333332\\
0.09  0.14270833333333333\\
0.08  0.12383814102564103\\
0.07  0.10482638888888889\\
0.06  0.081953125\\
0.05  0.0644375\\
0.04  0.04826388888888889\\
0.03  0.031614583333333335\\
0.02  0.01936085390946502\\
0.01  0.008054123711340205\\
};
\addlegendentry{$m=6$, $\epsilon_0=0.001$}

 \addplot [color=A5!70, dashed, line width=1.5pt, mark=square, mark options={solid, A5!70, mark size=2pt}]  table[row sep=crcr]{%
0.1  0.15506628787878787\\
0.09  0.13506944444444444\\
0.08  0.1104861111111111\\
0.07  0.09322916666666667\\
0.06  0.07272727272727272\\
0.05  0.05454382183908046\\
0.04  0.03954427083333333\\
0.03  0.02455929487179487\\
0.02  0.01329007768361582\\
0.01  0.005071804207119741\\
};
\addlegendentry{$m=6$, $\epsilon_0=0.1$}

 \addplot [color=A5, line width=1.5pt, mark=o, mark options={solid, A5, mark size=2pt}]
  table[row sep=crcr]{%
0.1  0.11450892857142857\\
0.09  0.09830729166666667\\
0.08  0.08065104166666667\\
0.07  0.0628125\\
0.06  0.04690563725490196\\
0.05  0.03553503787878788\\
0.04  0.02246279761904762\\
0.03  0.013130252100840336\\
0.02  0.00586376404494382\\
0.01  0.0015547340954274354\\
};
\addlegendentry{$m_{\text{oc}}=14$, $\epsilon_0=0.1$}

\end{axis}
\end{tikzpicture}  
    \vspace{-2ex}
    \caption{BP decoding results for the [[7,1,3]] CSS code with original and overcomplete check matrix ($L=32$).}\vspace*{\lsbelowcaption}
    \label{fig:FER71}
\end{figure}
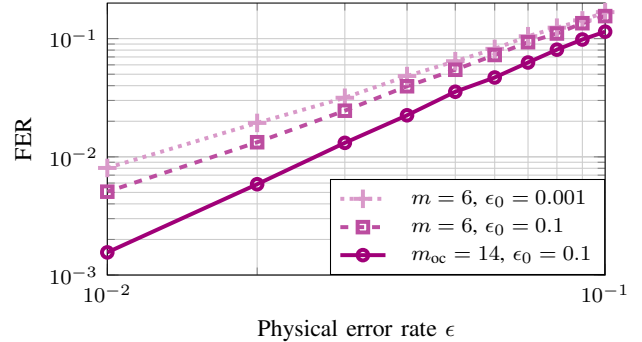

In this example, the decoder wrongly estimates the error because all CNs have a syndrome of $1$. Therefore, there is a strong indication that the error has a high weight. Except for \acp{VN} $\mathsf{v}_1$, $\mathsf{v}_2$, and $\mathsf{v}_4$ with degree 2, all VNs that are connected to more than 2 \acp{CN} are erroneously estimated in the first hard-decision step. Duplicating some rows of the check matrix does not help in this case\footnote{However, note that this problem could be solved by initializing the decoder with a very small $\epsilon_0$ such as $0.001$ at the cost of degrading the overall decoding performance (depicted in Fig.~\ref{fig:FER71}), as in most cases, the decoder has a tendency towards trivial errors.}.

Now let us use the overcomplete $\vec{H}_X$ and $\vec{H}_Z$ with 7 rows, i.e., we take all the linear combinations of the $3$ rows of the original $\vec{H}_{\text{BCH}}$ except the trivial case. The new syndrome is
\[
\vec{z}_{\text{oc}} = \begin{pmatrix}
1\;   1 \;  0 \;  1  \; 0\;  0 \;  1 \;  1 \;  1 \;  0 \;  1  \; 0 \;  0 \;  1\\
\end{pmatrix}.
\]
Although $\vec{z}_{\text{oc}}$ contains the same amount of information as $\vec{z}$ with respect to the coset where the error $\vec{\mathcal{E}}$ is located, the former is easier for the decoder to interpret. In the first hard-decision step, the error is correctly estimated. Fig.~\ref{fig:FER71} depicts the \ac{FER} curves corresponding to decoding using the overcomplete check matrix and the original check matrix for different $\epsilon_0$.

\vspace{-1ex}
\section{Neural Belief Propagation}
\label{sec:NBP}

In our \ac{NBP} decoder, additional weights are introduced for each $\vec{\Lambda}_i$ and each \ac{CN}. The update rules \eqref{eq:CNupdate} and \eqref{eq:VNupdate} are modified to
\begin{equation}
\Delta_{i\lto j} = (-1)^{z_j}\cdot 2w_{\mathsf{c},j}^{(\ell)}\tanh^{-1} \left(\prod_{i'\in \mathcal{N}(j)\backslash{i}}\tanh \frac{\lambda_{i'\to j}}{2}\right)
\end{equation}
and
\begin{equation}
\Gamma^{(\zeta)}_{i\to j} = w_{\mathsf{v},i}^{(\ell)}\Lambda_i^{(\zeta)} + \sum_{\substack{j'\in \mathcal{M}(i)\backslash{j}, \\ \langle \zeta, S_{j'i}\rangle=1}}\Delta_{i\lto j'},
\end{equation}
where $\ell$ is the index of the decoding iteration and $w_{\mathsf{c},j}^{(\ell)}$ and $w_{\mathsf{v},i}^{(\ell)}$ denote the trainable weights. The model is relatively simple compared to conventional \ac{NBP}, where weights are applied on each edge of the Tanner graph. For the codes considered in this work, using more model parameters did not improve the performance.

In \cite{liu2019neural}, a loss function for BP2 decoding has been proposed which takes degeneracy into account.
We extend this loss function to the BP4 case. Using $\vec{\Gamma}_{i}$ calculated by \eqref{eq:hard-decision} in the hard-decision step, we now calculate
\begin{equation}
\label{eq:loss_aid}
P(\langle \hat{e}_i, \eta \rangle=1|\vec{z}) = \left(1+\me^{-\lambda_{\eta}(\vec{\Gamma}_i)}\right)^{-1}
\end{equation}
for $\eta \in$ GF(4)$\backslash \{0\}$, indicating the estimated probability of the $i$-th error commuting with $\vec{X}$,  $\vec{Z}$, and $\vec{Y}$, respectively. Then proposed loss function can be written as
\begin{equation}
\label{eq:loss}
\mathcal{L}(\vec{\Gamma};\vec{e}) = \sum_{j=1}^{2n-m} f\left( \sum_{i=1}^{n} P\left(\langle e_i+\hat{e}_i, S_{ji}^{\perp} \rangle =1|\vec{z}\right)\right)
\end{equation}
where $f(x)=|\sin(\pi x/2)|$, as in \cite{liu2019neural}. The loss is summed up over all rows $\bm{S}^{\perp}_j$ of $\bm{S}^{\perp}$. For each row $j$, we sum up the values $P\left(\langle e_i+\hat{e}_i, S_{ji}^{\perp} \rangle =1|\vec{z}\right)$ for all the elements $S_{ji}^{\perp}$ in $\bm{S}^{\perp}_j$, representing the probability of $S_{ji}^{\perp}$ being unsatisfied after estimating $e_i$ as $\hat{e}_i$. It can be calculated as $P\left(\langle \hat{e}_i, S_{ji}^{\perp} \rangle =1 +\langle e_i,  S_{ji}^{\perp} \rangle|\vec{z}\right)$. When $\langle e_i,  S_{ji}^{\perp} \rangle =0$, we directly calculate $P\left(\langle \hat{e}_i, S_{ji}^{\perp} \rangle =1|\vec{z}\right)$ using \eqref{eq:loss_aid}, with $\eta$ being $S_{ji}^{\perp}$. When $\langle e_i,  S_{ji}^{\perp} \rangle =1$, we calculate $P\left(\langle \hat{e}_i, S_{ji}^{\perp} \rangle =0|\vec{z}\right)=1-P\left(\langle \hat{e}_i, S_{ji}^{\perp} \rangle =1|\vec{z}\right)$. As $f(x)$ approaches $0$ with $x$ approaching any even number, the loss function \eqref{eq:loss} is minimized when \eqref{eq:correction} holds.

During training, we use multi-loss~\cite{nachmani2016learning,nachmani2018deep} calculated as the average value of the loss function after every iteration until the loss is minimized. This helps to increase the magnitude of gradients corresponding to earlier decoding iterations.

\section{Numerical Results}
We assess the performance of the proposed decoder using Monte Carlo simulations. To ensure sufficiently accurate results, 300 frame errors are collected for each data point. We always use the refined BP4 decoder with a flooding schedule. The initial $\epsilon_0$ is always set to $0.1$, which yields a good decoding performance for the codes considered in this work. This improves the decoding performance when the actual physical error probability $\epsilon$ is too small~\cite{lai2021log}. Using a fixed initialization also avoids estimating the channel statistics, which is not always feasible.

We consider \ac{GB} codes proposed in~\cite{panteleev2021degenerate}. They are constructed by
$\vec{H}_X' = \begin{pmatrix}
\vec{A}&\vec{B}\\
\end{pmatrix}$ and $\vec{H}_Z' = \begin{pmatrix}
\trans{\vec{B}}&\trans{\vec{A}}\\
\end{pmatrix}$
where $\vec{A}$ and $\vec{B}$ are $\frac{n}{2}\times \frac{n}{2}$ square circulant matrices. For constructing the check matrix, $m/2$ rows are chosen from both $\vec{A}$ or $\vec{B}$. However, all the rows of $\vec{A}$ or $\vec{B}$ are linearly dependent. Therefore, the \ac{GB} codes naturally have an overcomplete set of checks. Further redundant rows are generated following the method described in Sec.~\ref{sec:soc}.
\subsection{\ac{NBP} Results}
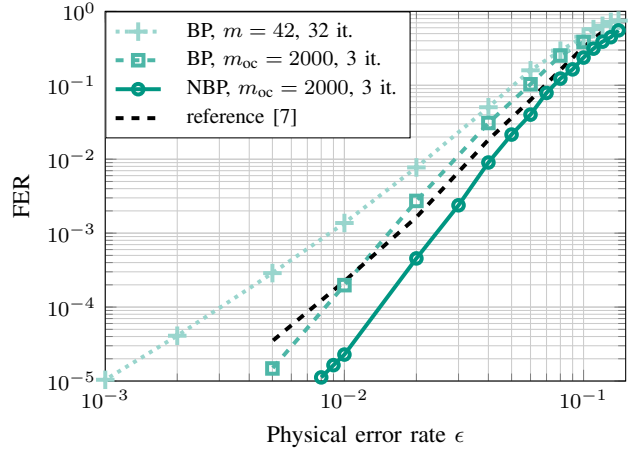
\begin{figure}[t]
     \begin{tikzpicture}
       \pgfplotsset{grid style={gray!40}}
       \pgfplotsset{every tick label/.append style={font=\footnotesize}}
        \begin{axis}[%
        name=ax1,
        scale only axis=false,
        width=8.5cm,
        height=6.5cm,
        xmin=0.001,
        xmax=0.15,
        xmode=log,
        ymode=log,
        ymin=1e-5,
        ymax=1,
        yminorticks,
        xminorticks,
        axis background/.style={fill=white, mark size=1.5pt},
        xmajorgrids,
        xminorgrids,
        ymajorgrids,
        yminorgrids,
        ylabel={FER},
                xlabel={Physical error rate $\epsilon$},
        label style={font=\small},
        legend cell align={left},
        legend style={at={(0,1)},anchor=north west,fill opacity=1, text opacity = 1,legend columns=1, row sep = 0pt,font=\footnotesize}     
]

\addlegendentry{BP, $m=42$, 32 it.}
\addlegendentry{BP, $m_{\text{oc}}=2000$, 3 it.}
\addlegendentry{NBP, $m_{\text{oc}}=2000$, 3 it.}

  \addplot [color=A3!40, dotted, line width=1.5pt, mark=+, mark options={solid, A3!40, fill=white, mark size=3.5pt}]table[row sep=crcr]{%
0.14  0.74921875\\
0.13  0.71953125\\
0.12  0.6375\\
0.11  0.575\\
0.1  0.45390625\\
0.08  0.29375\\
0.06  0.159375\\
0.04  0.05046875\\
0.02  0.007736895161290323\\
0.01  0.0013717296511627907\\
0.005  0.00028818167892156863\\
0.002  4.0986884197056945e-05\\
0.001  1.0396673064619321e-05\\
 };

 \addplot [color=A3!70, dashed, line width=1.5pt, mark=square, mark options={solid, A3!70, mark size=2pt}]
  table[row sep=crcr]{%
0.1  0.3848958333333333\\
0.08  0.2526041666666667\\
0.06  0.103125\\
0.04  0.03090277777777778\\
0.02  0.0027298850574712643\\
0.01  0.00019803548795944233\\
0.005  1.4795265515035188e-05\\
};

 \addplot [color=A3, line width=1.5pt, mark=o, mark options={solid, A3, mark size=2pt}]
  table[row sep=crcr]{%
0.14  0.553125\\
0.13  0.4473958333333333\\
0.12  0.3890625\\
0.11  0.31302083333333336\\
0.1  0.23697916666666666\\
0.09  0.1640625\\
0.08  0.12317708333333334\\
0.07  0.07881944444444444\\
0.06  0.04010416666666667\\
0.05  0.021549479166666666\\
0.04  0.009027777777777777\\
0.03  0.002383207070707071\\
0.02  0.00045553935860058307\\
0.01  2.279270633397313e-05\\
0.009  1.6394753678822777e-05\\
0.008  1.1196417146513115e-05\\
};

   \addplot [color=black, dashed, line width=1.5pt,mark=none ]table[x=P,y=WER,row sep=crcr]{
  P       BER        WER        FAR     numWords 
0.14 0.424051    0.632911   0.632911   158       \\     
0.12 0.32716     0.529101   0.529101   189       \\     
0.1  0.197368    0.328947   0.328947   304         \\   
0.08 0.0973571   0.161812   0.161812   618       \\     
0.06 0.0381043   0.0633312  0.0633312  1579        \\   
0.04 0.00974659  0.0182749  0.0182749  5472       \\    
0.02 0.000791928 0.00164414 0.00164414 60822     \\     
0.015 0.000327151 0.000721657 0.000721657 138570   \\      
0.01  9.84289e-05 0.000223702 0.000223702 447023 \\       
0.005 1.26819e-05 3.47448e-05 3.47448e-05 2878128  \\   
 };
\addlegendentry{reference~\cite{panteleev2021degenerate}}

\end{axis}
\end{tikzpicture}  
    \vspace{-2ex}
    \caption{FER vs. depolarizing probability curves for the [[48,6,8]] code}\vspace*{\lsbelowcaption}\label{fig:FER48}
\end{figure}
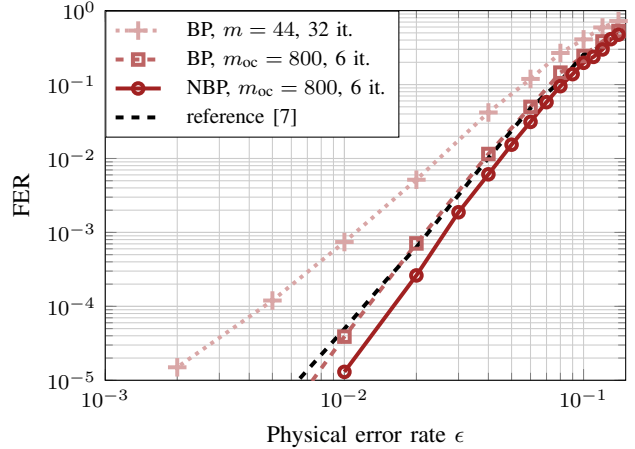
\begin{figure}[t]
    \vspace{-4ex}
     \begin{tikzpicture}
       \pgfplotsset{grid style={gray!40}}
       \pgfplotsset{every tick label/.append style={font=\footnotesize}}
        \begin{axis}[%
        name=ax1,
        width=8.5cm,
        height=6.5cm,
        xmin=0.001,
        xmax=0.15,
        xmode=log,
        ymode=log,
        ymin=1e-5,
        ymax=1,
        yminorticks,
        axis background/.style={fill=white, mark size=1.5pt},
        xmajorgrids,
        xminorgrids,
        ymajorgrids,
        yminorgrids,
        ylabel={FER},
                xlabel={Physical error rate $\epsilon$},
        label style={font=\small},
        legend cell align={left},
        legend style={at={(0,1)},anchor=north west,fill opacity=1, text opacity = 1,legend columns=1, row sep = 0pt,font=\footnotesize}
]

\addlegendentry{BP, $m=44$, 32 it.}
\addlegendentry{BP, $m_{\text{oc}}=800$, 6 it.}
\addlegendentry{NBP, $m_{\text{oc}}=800$, 6 it.}

 \addplot[color=A4!40, dotted, line width=1.5pt, mark=+, mark options={solid, A4!40, fill=white, mark size=3.5pt}]
  table[row sep=crcr]{%
0.14  0.72265625\\
0.12  0.5921875\\
0.1  0.40859375\\
0.08  0.2671875\\
0.06  0.119140625\\
0.04  0.04192708333333333\\
0.02  0.005163043478260869\\
0.01  0.0007418178233438486\\
0.005  0.00011994626407369498\\
0.002  1.4997120552853852e-05\\
};

 \addplot [color=A4!70, dashed, line width=1.5pt, mark=square, mark options={solid, A4!70, mark size=2pt}]
  table[row sep=crcr]{%
0.14  0.53046875\\
0.12  0.3828125\\
0.1  0.24375\\
0.08  0.144921875\\
0.06  0.04984375\\
0.04  0.011495535714285715\\
0.02  0.0007102272727272727\\
0.01  3.899251343742002e-05\\
0.005  1.7996544663424623e-06\\
};

 \addplot [color=A4, line width=1.5pt, mark=o, mark options={solid, A4, mark size=2pt}]
  table[row sep=crcr]{%
0.14  0.47265625\\
0.13  0.4109375\\
0.12  0.2984375\\
0.11  0.23671875\\
0.1  0.196875\\
0.09  0.13828125\\
0.08  0.09505208333333333\\
0.07  0.0578125\\
0.06  0.03125\\
0.05  0.015380859375\\
0.04  0.006109775641025641\\
0.03  0.001875\\
0.02  0.00026216442953020134\\
0.01  1.2997504479140006e-05\\
};

   \addplot [color=black, dashed, line width=1.5pt,mark=none ]table[x=P,y=WER,row sep=crcr]{
  P      BER         WER         FAR     numWords 
0.14 0.373134   0.497512    0.497512    201      \\       
0.12 0.239057   0.3367      0.3367      297       \\      
0.1  0.185751   0.254453    0.254453    393       \\      
0.08 0.0860335  0.111732    0.111732    895       \\      
0.06 0.0351171  0.0477783   0.0477783   2093       \\     
0.04 0.00698805 0.0101276   0.0101276   9874      \\      
0.02 0.00042262 0.000630776 0.000630776 158535    \\      
0.015 0.000155857 0.000229202 0.000229202 436297    \\      
0.01  3.46871e-05 4.9553e-05  4.9553e-05  2018043   \\      
0.005 2.63069e-06 4.09219e-06 4.09219e-06 11974032 \\
 };
\addlegendentry{reference~\cite{panteleev2021degenerate}}

\end{axis}
\end{tikzpicture}  
    \vspace{-2ex}
    \caption{FER vs. depolarizing probability curves for the [[46,2,9]] code}\vspace*{\lsbelowcaption}\label{fig:FER46}
\end{figure}
Figure~\ref{fig:FER48} and Fig.~\ref{fig:FER46} show the simulation results for the [[48,6,8]] and [[46,2,9]] \ac{GB} codes (codes A3 and A4 in~\cite{panteleev2021degenerate}). We compare the decoding performance for the two codes using refined \ac{BP} decoding with different settings. For BP4 decoding over the original Tanner graph, 32 iterations are performed. For the [[48,6,8]] code, we construct the overcomplete check matrix using $48$ rows of weight $8$ and $1952$ rows of weight $12$ (no checks of weight $10$ were found). For the [[46,2,9]] code, we use $46$ rows of weight $8$ and $754$ rows of weight $10$. We observe that in this case, the number of required decoding iterations can be greatly reduced. Only 3 iterations are necessary for the [[48,6,8]] code and 6 iterations for the [[46,2,9]] code. Further increasing the number of iterations does not improve the performance noticeably. We compare the \ac{FER} results with the reference results taken from~\cite{panteleev2021degenerate} where normalized min-sum decoding with layered scheduling is followed by an \ac{OSD} post-processing step. The reference outperforms the original BP decoding which suffers from a higher error floor. However, \ac{BP} decoding using the overcomplete check matrix performs comparably to the reference.

\begin{figure}[t]
     \begin{tikzpicture}
       \pgfplotsset{grid style={gray!40}}
       \pgfplotsset{every tick label/.append style={font=\footnotesize}}
        \begin{axis}[%
        name=ax1,
        width=8.5cm,
        height=5.5cm,
        xmin=1,
        xmax=10,
        ymin=0,
        ymax=1,
        yminorticks,
        axis background/.style={fill=white, mark size=1.5pt},
        xmajorgrids,
        xminorgrids,
        ymajorgrids,
        yminorgrids,
        ytick={0,0.1,...,1},
        ylabel={FER},
        xlabel={Error weight},
        label style={font=\small},
        legend cell align={left},
        legend style={at={(0,1)},anchor=north west,fill opacity=1, text opacity = 1,legend columns=1, row sep = 0pt,font=\footnotesize}
]
\addlegendentry{BP, $m=42$, 32 it.}
\addlegendentry{BP, $m_{\text{oc}}=2000$, 3 it.}
\addlegendentry{NBP, $m_{\text{oc}}=2000$, 3 it.}
\addplot [color=A3!40, dotted, line width=1.5pt, mark=+, mark options={solid, A3!40, fill=white, mark size=3.5pt}]
  table[row sep=crcr]{
1  0.0\\
2  0.008289930555555556\\
3  0.043619791666666664\\
4  0.166796875\\
5  0.39921875\\
6  0.77734375\\
7  0.95859375\\
8  0.98125\\
9  0.99765625\\
10  1.0\\
11  1.0\\
12  1.0\\
13  1.0\\
};

 \addplot [color=A3!70, dashed, line width=1.5pt, mark=square, mark options={solid, A3!70, mark size=2pt}]
  table[row sep=crcr]{%
1  0.0\\
2  0.0\\
3  9.765625e-05\\
4  0.10052083333333334\\
5  0.3703125\\
6  0.65859375\\
7  0.8890625\\
8  0.9765625\\
9  0.99453125\\
10  0.9984375\\
11  0.99921875\\
12  1.0\\
13  1.0\\
};

 \addplot [color=A3, line width=1.5pt, mark=o, mark options={solid, A3, mark size=2pt}]
  table[row sep=crcr]{%
1  0.0\\
2  0.0\\
3  0.0\\
4  0.008081896551724138\\
5  0.066015625\\
6  0.2828125\\
7  0.5578125\\
8  0.8265625\\
9  0.94140625\\
10  0.9859375\\
11  0.99453125\\
12  0.9984375\\
13  1.0\\
};

\end{axis}
\end{tikzpicture}  
    \vspace{-2ex}
    \caption{FER vs. error weight curves for the [[48,6,8]] code}\vspace*{\lsbelowcaption}\label{fig:FER48w}
\end{figure}
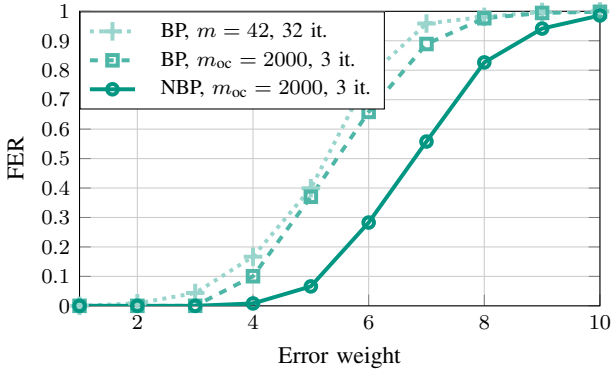

\begin{figure}[t]
    \vspace{-3ex}
    \begin{tikzpicture}
       \pgfplotsset{grid style={gray!40}}
       \pgfplotsset{every tick label/.append style={font=\footnotesize}}
        \begin{axis}[%
        name=ax1,
        width=8.5cm,
        height=5.5cm,
        xmin=1,
        xmax=10,
        ymin=0,
        ymax=1,
        yminorticks,
        axis background/.style={fill=white, mark size=1.5pt},
        xmajorgrids,
        xminorgrids,
        ymajorgrids,
        yminorgrids,
        ytick={0,0.1,...,1},
        ylabel={FER},
        xlabel={Error weight},
        label style={font=\small},
        legend cell align={left},
        legend style={at={(0,1)},anchor=north west,fill opacity=1, text opacity = 1,legend columns=1, row sep = 0pt,font=\footnotesize}
]
\addlegendentry{BP, $m=44$, 32 it.}
\addlegendentry{BP, $m_{\text{oc}}=800$, 6 it.}
\addlegendentry{NBP, $m_{\text{oc}}=800$, 6 it.}
  \addplot [color=A4!40, dotted, line width=1.5pt, mark=+, mark options={solid, A4!40, fill=white, mark size=3.5pt}]
  table[row sep=crcr]{%
1  0.0\\
2  0.00625\\
3  0.02958984375\\
4  0.13359375\\
5  0.4359375\\
6  0.790625\\
7  0.9609375\\
8  0.9890625\\
9  1.0\\
10  1.0\\
};

 \addplot [color=A4!70, dashed, line width=1.5pt, mark=square, mark options={solid, A4!70, mark size=2pt}]
  table[row sep=crcr]{
1  0.0\\
2  0.0\\
3  0.0\\
4  0.028125\\
5  0.171484375\\
6  0.4375\\
7  0.70859375\\
8  0.87734375\\
9  0.95546875\\
10  0.97890625\\
};

 \addplot [color=A4, line width=1.5pt, mark=o, mark options={solid, A4, mark size=2pt}]
  table[row sep=crcr]{%
1  0.0\\
2  0.0\\
3  0.0\\
4  0.00171875\\
5  0.066015625\\
6  0.26875\\
7  0.53515625\\
8  0.7765625\\
9  0.8984375\\
10  0.95859375\\
11  0.97578125\\
12  0.9890625\\
13  0.9875\\
};

\end{axis}
\end{tikzpicture}  
    \vspace{-2ex}
    \caption{FER vs. error weight curves for the [[46,2,9]] code}\vspace*{\lsbelowcaption}\label{fig:FER46w}     
\end{figure}
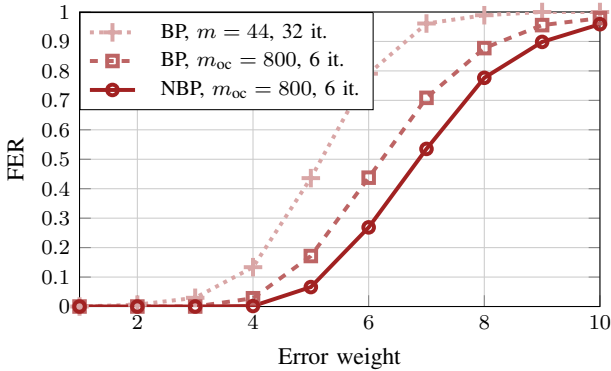

The \acp{FER} can be further reduced by \ac{NBP}. Using the loss function described in Sec.~\ref{sec:NBP}, the weights $w_{\mathsf{c,j}}^{(\ell)}$ and $w_{\mathsf{v},i}^{(\ell)}$ are optimized using gradient descent. During training, the learning rate is $0.001$ and the Adam optimizer~\cite{kingma2015adam} is used. We first pre-train the model on $1500$ batches of batch size $100$ consisting of random weight $2$ and weight $3$ errors. Then we train the model on $600$ batches of batch size $100$ consisting of random weight $3$ to $9$ errors. Weight $1$ errors are omitted for both codes, as the loss value of weight $1$ errors is always almost zero. This training approach is based on our observation that directly training the NN using random errors causes the loss to diverge. The conjectured reason is that too many uncorrectable errors increase the gradient noise which makes the model prone to be stuck in a local minimum. We observe that on average, pre-training yields higher weights on low-degree CNs, which is consistent with our experience that low-degree checks are more helpful in decoding. Subsequent training on high-weight errors further improves the decoding performance. After training, our \ac{NBP} decoder outperforms the reference results, with the additional advantage of a very small number of required decoding iterations. 

To show that the performance gain is not simply due to overfitting, i.e., the decoder simply remembering how to decode all errors of weight 2 and weight 3, we plot the \ac{FER} results when decoding errors of different weights in Fig.~\ref{fig:FER48w} and Fig.~\ref{fig:FER46w}. The figures show that the trained \ac{NBP} decoder also outperforms conventional \ac{BP} on high-weight errors.

\subsection{Effect of Check Matrix with Redundant Rows}
We show that adding redundant rows is often beneficial for improving the decoding performance. However, the added rows need to be chosen carefully. We observe that the performance improves with a large number of redundant rows if the maximum CN degree stays unchanged.
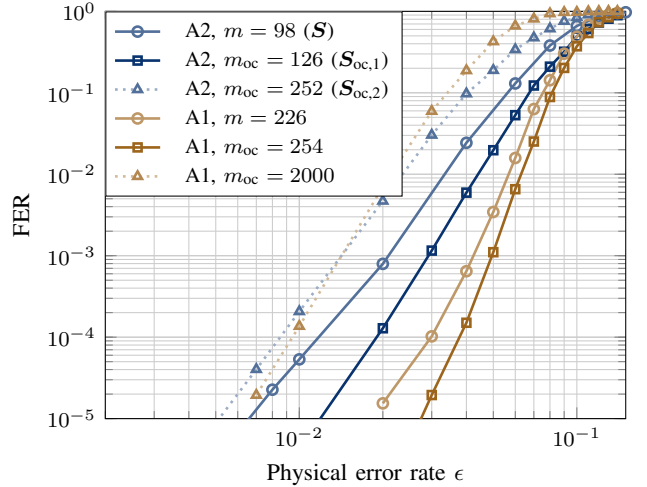
\begin{figure}[tb]
    \centering
     \begin{tikzpicture}
       \pgfplotsset{grid style={gray!40}}
       \pgfplotsset{every tick label/.append style={font=\footnotesize}}
        \begin{axis}[%
        name=ax1,
        width=8.5cm,
        height=7cm,
        xmin=0.002,
        xmax=0.15,
        xmode=log,
        ymode=log,
        ymin=1e-5,
        ymax=1,
        yminorticks,
        axis background/.style={fill=white, mark size=1.5pt},
        xmajorgrids,
        xminorgrids,
        ymajorgrids,
        yminorgrids,
        ylabel={FER},
                xlabel={Physical error rate $\epsilon$},
        label style={font=\small},
        legend cell align={left},
        legend style={at={(0,1)},anchor=north west,fill opacity=1, text opacity = 1,legend columns=1, row sep = 0pt,font=\footnotesize}
]

  \addplot [color=A2!70, line width=1pt,mark=o, mark options={solid,,A2!70, mark size=1.9pt}]table[row sep=crcr]{
0.15  0.9739583333333334\\
0.12  0.8614583333333333\\
0.1  0.6619791666666667\\
0.08  0.3807291666666667\\
0.06  0.13020833333333334\\
0.04  0.024330357142857143\\
0.02  0.0007944023569023569\\
0.01  5.338291746641075e-05\\
0.008 2.26461e-05\\
0.006 0.000006733\\
 };
\addlegendentry{A2, $m=98$ ($\vec{S}$)}

  \addplot [color=A2, line width=1pt,mark=square, mark options={solid, A2, mark size=1.5pt}]table[row sep=crcr]{
0.14  0.8973958333333333\\
0.13  0.8125\\
0.12  0.7296875\\
0.11  0.6072916666666667\\
0.1  0.49114583333333334\\
0.09  0.31927083333333334\\
0.08  0.209375\\
0.07  0.12291666666666666\\
0.06  0.053125\\
0.05  0.0197265625\\
0.04  0.005945362\\
0.03  0.0011557177615571777\\
0.02  0.0001286008230452675\\
0.01  0.00000424\\
 };
\addlegendentry{A2, $m_{\text{oc}}=126$ ($\vec{S}_{\text{oc,1}}$)}

  \addplot [color=A2!40, dotted, line width=1pt,mark=triangle, mark options={solid,,A2!70, mark size=1.9pt}]table[row sep=crcr]{
0.14 0.972056\\
0.13 0.973948\\
0.12 0.954902\\
0.11 0.907721\\
0.1 0.842014\\
0.09 0.756289\\
0.08 0.612484\\
0.07 0.474268\\
0.06 0.339578\\
0.05 0.188843\\
0.04 0.0974555\\
0.03 0.0302947\\
0.02 0.00469441\\
0.01 0.000206423\\
0.007 4.02166e-05\\
0.005 0.0000088\\
 };
\addlegendentry{A2, $m_{\text{oc}}=252$  ($\vec{S}_{\text{oc,2}}$)}

  \addplot [color=A1!60, line width=1pt,mark=o, mark options={solid, A1!60, mark size=1.9pt}]table[row sep=crcr]{
0.14  0.9552083333333333\\
0.13  0.9036458333333334\\
0.12  0.8192708333333333\\
0.11  0.6822916666666666\\
0.1  0.4875\\
0.09  0.3020833333333333\\
0.08  0.14270833333333333\\
0.07  0.06319444444444444\\
0.06  0.015833333333333335\\
0.05  0.0034420289855072463\\
0.04  0.0006451474622770919\\
0.03  0.00010225785340314136\\
0.02  1.5395073576455534e-05\\
 };
\addlegendentry{A1, $m=226$}
  \addplot [color=A1!90, line width=1pt,mark=square, mark options={solid, A1!90, mark size=1.5pt}]table[row sep=crcr]{
0.14  0.9375\\
0.13  0.8453125\\
0.12  0.7359375\\
0.11  0.5416666666666666\\
0.1  0.3723958333333333\\
0.09  0.20208333333333334\\
0.08  0.08880208333333334\\
0.07  0.025223214285714286\\
0.06  0.006553819444444445\\
0.05  0.0011040199530516432\\
0.04  0.00015024038461538462\\
0.03 0.000019477\\
0.027  0.000008799\\
 };
\addlegendentry{A1, $m_{\text{oc}}=254$}

  \addplot [color=A1!40, dotted, line width=1pt,mark=triangle, mark options={solid,,A1!70, mark size=1.9pt}]table[row sep=crcr]{
0.14 1\\
0.13 1\\
0.12 0.997963\\
0.11 0.993865\\
0.1 0.97561\\
0.09 0.971545\\
0.08 0.9375\\
0.07 0.808547\\
0.06 0.662841\\
0.05 0.425388\\
0.04 0.187587\\
0.03 0.059401\\
0.02 0.00700557\\
0.01 0.000136245\\
0.007 1.93998e-05\\
 };
\addlegendentry{A1, $m_{\text{oc}}=2000$}

\end{axis}
\end{tikzpicture}  
    \vspace{-2ex}
    \caption{FER vs. depolarizing probability $\epsilon$ curves for the $[[126,28,8]]$ (A2) and $[[254,28,d]]$ (A1) GB code with original and overcomplete check matrix ($L=32$).}
    \vspace*{\lsbelowcaption}
    \label{fig:FER126}
\end{figure}
We compare the \acp{FER} of two \ac{GB} codes with parameters $[[126,28,8]]$ (code A2 in~\cite{panteleev2021degenerate}) and $[[254,28,d]]$ (code A1 in~\cite{panteleev2021degenerate}) when decoding with different check matrices. Checks of low weight are always added first when constructing the overcomplete check matrix.

For the $[[126,28,8]]$ code, we consider three check matrices: The original check matrix $\vec{S}$ with $98$ rows of weight $10$, an overcomplete check matrix $\vec{S}_{\text{oc,1}}$ with $126$ rows of weight $10$, and another overcomplete check matrix $\vec{S}_{\text{oc,2}}$ consisting of $\vec{S}_{\text{oc,1}}$ and $126$ rows of weight $16$. Decoding with $\vec{S}_{\text{oc,1}}$ outperforms the decoder based on $\vec{S}$ by nearly an order of magnitude at $\epsilon=0.02$. However, decoding with $\vec{S}_{\text{oc,2}}$ shows a performance degradation.
For the $[[254,28]]$ code, similar results can be observed. Compared with the results of decoding with the original check matrix with $226$ rows, adding rows of weight $10$ ($m_{\text{oc}}=254$) improves the decoding performance. However, further adding rows of weight $18$ ($m_{\text{oc}}=2000$) degrades the decoding performance.

\section{Conclusion}
In this paper, \ac{BP}4 decoding for \ac{QLDPC} codes based on a check matrix with redundant rows is proposed and investigated. The method is shown to be effective for several \ac{QLDPC} codes in improving the decoding results. Moreover, combined with \ac{NBP}, the performance can be further improved. As a large number of node updates are performed in parallel, which reduces the number of required decoding iterations, the decoder has a very small decoding latency. Moreover, it is well-accepted that the BP decoding performance is impacted by the structure of the underlying Tanner graph. Due to the symplectic criterion, optimizing the structure of the Tanner graph for QLDPC codes is more difficult than for classical LDPC codes. Using a check matrix with redundant rows gives us some degree of freedom in optimizing the graph structure.


\begin{thebibliography}{00}
    \bibitem{tillich2013quantum}
    J.-P. Tillich and G.~Z{\'e}mor, ``Quantum {LDPC} codes with positive rate and minimum distance proportional to the square root of the blocklength,''
    \emph{IEEE Transactions on Information Theory}, vol.~60, no.~2, pp.
    1193--1202, 2013.
    
    \bibitem{panteleev2021quantum}
    P.~Panteleev and G.~Kalachev, ``Quantum {LDPC} codes with almost linear minimum
    distance,'' \emph{IEEE Transactions on Information Theory}, vol.~68, no.~1,
    pp. 213--229, 2021.
    
    \bibitem{panteleev2022asymptotically}
    ------, ``Asymptotically good quantum and locally testable classical {LDPC} codes,'' Preprint, available online at https://arxiv.org/abs/2111.03654.
    
    \bibitem{breuckmann21balanced}
    N.~P.~Breuckmann, and J.~N.~Eberhardt. ``Balanced product quantum codes,'' \emph{IEEE Transactions on Information Theory}, vol.~67, no.~10, pp.~6653--6674, 2021.
    
    \bibitem{kuo2020refined}
    K.-Y. Kuo and C.-Y. Lai, ``Refined belief propagation decoding of sparse-graph quantum codes,'' \emph{IEEE Journal on Selected Areas in Information Theory}, vol.~1, no.~2, pp. 487--498, 2020.
    
    \bibitem{lai2021log}
    C.-Y. Lai and K.-Y. Kuo, ``Log-domain decoding of quantum {LDPC} codes over binary finite fields,'' \emph{IEEE Transactions on Quantum Engineering}, vol.~2, pp. 1--15, 2021.
    
    \bibitem{panteleev2021degenerate}
    P.~Panteleev and G.~Kalachev, ``Degenerate quantum {LDPC} codes with good finite length performance,'' \emph{Quantum}, vol.~5, 2021.
    
    \bibitem{raveendran2021trapping}
    N.~Raveendran and B.~Vasi{\'c}, ``Trapping sets of quantum {LDPC} codes,'' \emph{Quantum}, vol.~5, pp.~1--19, 2021.
    
    \bibitem{rigby2019modified}
    A.~Rigby, J.~C.~Olivier, and P.~ Jarvis, ``Modified belief propagation decoders for quantum low-density parity-check codes,'' \emph{Physical Review A}, vol.~100, no.~1, 2019.
    
    
    
    
    \bibitem{roffe2020decoding}
    J.~Roffe, D.~R. White, S.~Burton, and E.~Campbell, ``Decoding across the
    quantum low-density parity-check code landscape,'' \emph{Physical Review
        Research}, vol.~2, no.~4, 2020.
    \bibitem{poulin2008iterative}
    D.~Poulin and Y.~Chung, ``On the iterative decoding of sparse quantum codes,'' \emph{Quantum Information and Computation}, vol.~8, no.~10, pp.~987--1000, 2008.
    
    \bibitem{wang2012enhancedfeedback}
    Y. Wang, B. C. Sanders, B. Bai, and X. Wang., ``Enhanced feedback iterative decoding of sparse quantum codes,'' \emph{IEEE Transactions on Information Theory}, vol.~58, no.~2, pp.~1231--1241, 2012.
    
    \bibitem{crest2022stabilizer}
    J.~du Crest, M.~Mhalla, and V.~Savin, ``Stabilizer inactivation for message-passing decoding of quantum {LDPC} codes,'' Preprint, available online at https://arxiv.org/abs/2205.06125.
    
    \bibitem{nachmani2016learning}
    E.~Nachmani, Y.~Be'ery, and D.~Burshtein, ``Learning to decode linear codes
    using deep learning,'' in \emph{Proc. Annual Allerton Conference on Communication, Control, and Computing}, pp. 341--346, 2016.
    
    \bibitem{nachmani2018deep}
    E.~Nachmani, E.~Marciano, L.~Lugosch, W.~J. Gross, D.~Burshtein, and Y.~Be’ery, ``Deep learning methods for improved decoding of linear codes,'' \emph{IEEE Journal of Selected Topics in Signal Processing}, vol.~12, no.~1, pp. 119--131, 2018.
    
    \bibitem{lian2019learned}
    M.~Lian, F.~Carpi, C.~H{\"a}ger, and H.~D. Pfister, ``Learned
    belief-propagation decoding with simple scaling and {SNR} adaptation,'' in
    \emph{Proc. IEEE International Symposium on Information Theory (ISIT)}, 2019.
    
    
    \bibitem{gottesman1997stabilizer}
    D.~Gottesman, \emph{Stabilizer codes and quantum error correction}, Ph.D. thesis, California Institute of Technology, 1997.
    
    \bibitem{Calderbank1998quantum}
    A.~R.~Calderbank, E.~M.~Rains, P.~W.~Shor, and N.~J.~A.~Sloane, ``Quantum error correction via codes over {GF}(4),'' \emph{{IEEE} Transactions on Information Theory}, vol.~44, no.~4, pp.~1369--1387, 1998.
    
    
    
    \bibitem{liu2019neural}
    T.~Liu and D.~Poulin, ``Neural belief-propagation decoders for quantum error-correcting codes,'' \emph{Physical review letters}, vol.~122, no.~20, 2019.
    
    
    
    
    \bibitem{xiao2019neural}
    X.~Xiao, N.~Raveendran, and B.~Vasi{\'c}, ``Neural-net decoding of quantum {LDPC} codes with straight-through estimators,'' in \emph{Proc. Information Theory and Applications Workshop (ITA)}, 2019.
    
    
    \bibitem{buchberger2020pruning}
    A.~Buchberger, C.~H{\"a}ger, H.~D.~Pfister, L.~Schmalen, and A.~Graell~i~Amat, ``Pruning and quantizing neural belief propagation decoders,'' \emph{{IEEE} Journal on Selected Areas in Communications}, vol.~39, no.~7, pp.~1957--1966, 2020.
    
    \bibitem{halford2006random}
    T.~R.~Halford and K.~M.~Chugg, ``Random redundant soft-in soft-out
    decoding of linear block codes,'' in \emph{Proc. IEEE International Symposium on Information Theory (ISIT)}, pp. 2230–2234, 2006.
    
    \bibitem{bossert1986hard}
    M.~Bossert and F.~Hergert, ``Hard- and soft-decision decoding beyond
    the half minimum distance—an algorithm for linear codes,'' \emph{IEEE Transactions on Information Theory}, vol. 32, no. 5, pp. 709–714, 1986.
    
    \bibitem{kothiyal2005iterative}
    A.~Kothiyal, O.~Y.~Takeshita, W.~Jin, and M.~Fossorier, ``Iterative reliability-based decoding of linear block codes with adaptive belief propagation,'' \emph{IEEE Communications Letters}, vol. 9, no. 12, pp. 1067–1069, 2005.
    
    \bibitem{Jiang2006iterative}
    J.~Jiang and K.~R.~Narayanan, ``Iterative soft-input soft-output decoding of Reed-Solomon codes by adapting the parity-check matrix,'' \emph{IEEE Transactions on Information Theory}, vol. 52, no. 8, pp. 3746–3756, 2006.
    
    
    \bibitem{Leon88}
    J.~S.~Leon, ``A probabilistic algorithm for computing minimum weights of large error-correcting codes,'' \emph{IEEE Transactions on Information Theory}, vol. 34, no. 5, pp. 1354-1359, 1988.
    
    \bibitem{kingma2015adam}
    D.~P.~Kingma and J.~L.~Ba, “Adam: A method for stochastic optimization,” in \emph{Proc. International Conference on Learning Representations (ICLR)}, 2015.
    
\end{thebibliography}
\end{document}